\documentclass[aps,pra,reprint,amsmath,superscriptaddress,noshowpacs,floatfix]{revtex4-1}

\usepackage{graphicx}
\usepackage{multirow}

\begin{document}

\title{Insights and challenges of applying the $GW$ method to transition metal oxides}

\author{Georgy Samsonidze}
\affiliation{Research and Technology Center, Robert Bosch LLC, Cambridge, Massachusetts 02142, USA}
\author{Cheol-Hwan Park}
\affiliation{Department of Physics, Seoul National University, Seoul 151-747, Korea}
\author{Boris Kozinsky}
\affiliation{Research and Technology Center, Robert Bosch LLC, Cambridge, Massachusetts 02142, USA}

\date{\today}

\begin{abstract}
The \textit{ab initio} $GW$ method is considered as the most accurate approach for calculating the band gaps 
of semiconductors and insulators. Yet its application to transition metal oxides (TMOs) has been hindered 
by the failure of traditional approximations developed for conventional semiconductors. In this work, 
we examine the effects of these approximations on the values of band gaps for ZnO, Cu$_2$O, and TiO$_2$. 
In particular, we explore the origin of the differences between the two widely used plasmon-pole models. 
Based on the comparison of our results with the experimental data and previously published calculations, 
we discuss which approximations are suitable for TMOs and why.
\\
\\
This is an author-created, un-copyedited version of an article published in 
Journal of Physics: Condensed Matter. IOP Publishing Ltd is not responsible 
for any errors or omissions in this version of the manuscript or any version 
derived from it. The Version of Record is available online at 
\href{http://dx.doi.org/10.1088/0953-8984/26/47/475501}{doi:10.1088/0953-8984/26/47/475501}.
\end{abstract}

%\pacs{71.10.-w, 71.15.Mb, 71.20.Nr}

\maketitle

\section{Introduction}
\label{sec:intro}

Many-body perturbation theory within the $GW$ approximation has been successfully used to 
describe the electronic spectra of $sp$-bonded semiconductors and insulators from first principles 
\cite{hedin-65-gw, hedin-70-gw, strinati-80-gw, strinati-82-gw, hybertsen-85-gw, godby-86-gw, 
hybertsen-86-gw, godby-88-gw, aryasetiawan-98-gw, onida-02-gw}. However, application of the 
$GW$ methodology to materials with localized $d$-electrons, such as transition metal oxides (TMOs), 
has revealed some controversial results. One of the heavily debated topics is the $GW$ band gap 
of ZnO for which values ranging from 2.1 to 3.9 eV have been reported \cite{usuda-02-zno-gw, 
kotani-07-zno-scgw, shishkin-07-zno-gw, fuchs-07-zno-gw, schleife-08-zno-gw, bechstedt-09-scgw, 
shih-10-zno-gw, friedrich-11-zno-gw, dixit-11-zno-gw, yan-11-zno-gw, stankovski-11-zno-gw, 
berger-12-zno-gw-eet, miglio-12-zno-gw, lany-13-tmo-gw, huser-13-zno-gw, larson-13-zno-ppm, klimes-14-zno-gw}. 
This wide variation can be attributed to the use of different self-consistent schemes 
\cite{van-schilfgaarde-06-scgw, bruneval-06-scgw, shishkin-07-zno-gw, fuchs-07-zno-gw, 
kotani-07-zno-scgw, bechstedt-09-scgw}, plasmon-pole models (PPMs) \cite{stankovski-11-zno-gw, 
miglio-12-zno-gw, larson-13-zno-ppm}, and starting points \cite{schleife-08-zno-gw, 
shih-10-zno-gw, yan-11-zno-gw}, as well as to a false convergence behavior as discussed 
in Ref.~\onlinecite{shih-10-zno-gw} and to the basis set convergence issues as discussed 
in Refs.~\onlinecite{friedrich-11-zno-gw, klimes-14-zno-gw}. At the same time, it is difficult to 
pinpoint the contributions of each approximation (self-consistent scheme, PPM, and starting point) 
to the total difference, since the different results reported in the literature were obtained 
with different codes and with different sets of numerical parameters.

The motivation behind the present study was to systematically 
isolate the contributions of these approximations. For that purpose 
we performed multiple $GW$ calculations for three TMOs (wurtzite ZnO, 
cuprite Cu$_2$O, and rutile TiO$_2$) using many possible combinations of these approximations. 
Analyzing the results of these calculations allowed us to collect valuable information about 
the validity and applicability of these approximations. We were able to 
show that the theoretically justified choice of approximations gives 
the best agreement with experiment for all the materials studied. 
We further discuss the origin of the differences 
between the two widely used PPMs, and we demonstrate how one of them can be 
modified to give better accuracy as compared to the results of higher level calculations.

The paper is organized as follows. Sec.~\ref{sec:theor} gives the theoretical background, 
followed by the computational details in Sec.~\ref{sec:comp}. Sec.~\ref{sec:res} presents 
the results and a discussion thereof. The main findings of this work are summarized in 
Sec.~\ref{sec:sum}.

\section{Theoretical background}
\label{sec:theor}

Within the $GW$ approximation, the electron self-energy operator $\Sigma$ 
is given by \cite{hedin-65-gw, hybertsen-86-gw, aryasetiawan-98-gw, 
onida-02-gw, deslippe-12-bgw, li-12-gw-ae}:
\begin{eqnarray}
\Sigma{(\mathbf{r}, \mathbf{r}'; \omega)} &=& 
\frac{i}{2\pi} \int d\omega' e^{i\omega'\eta} \nonumber \\
&& \times G{(\mathbf{r}, \mathbf{r}'; \omega + \omega')} 
W{(\mathbf{r}, \mathbf{r}'; \omega')}
\label{eqn:sigma}
\end{eqnarray}
where $\mathbf{r}$ is the spatial coordinate, $\omega$ is the energy, 
$\eta$ is a positive infinitesimal, $G$ is the Green's function, and 
$W$ is the screened Coulomb potential. The expression for $G$ is:
\begin{equation}
G{(\mathbf{r}, \mathbf{r}'; \omega)} = \sum_{n\mathbf{k}} \frac 
{\psi_{n\mathbf{k}}^\mathrm{QP}{(\mathbf{r})} {\psi_{n\mathbf{k}}^\mathrm{QP}}^\ast{(\mathbf{r}')}} 
{\omega - E_{n\mathbf{k}}^\mathrm{QP} - i \eta_{n\mathbf{k}}}
\label{eqn:g}
\end{equation}
where $n$ is the band index, $\mathbf{k}$ is the Bloch wave vector, 
$\psi_{n\mathbf{k}}^\mathrm{QP}{(\mathbf{r})}$ is the quasiparticle orbital, 
$E_{n\mathbf{k}}^\mathrm{QP}$ is the quasiparticle energy, and 
$\eta_{n\mathbf{k}}$ is a positive (negative) infinitesimal for 
occupied (unoccupied) states. The expression for $W$ is:
\begin{equation}
W{(\mathbf{r}, \mathbf{r}'; \omega)} = \int d\mathbf{r}'' 
\epsilon^{-1}{(\mathbf{r}, \mathbf{r}''; \omega)} v{(\mathbf{r}'' - \mathbf{r}')}
\label{eqn:w}
\end{equation}
where $\epsilon$ is the microscopic dielectric function, 
$v{(\mathbf{r})} = e^2 / \left|\mathbf{r}\right|$ is the bare Coulomb potential, 
and $e$ is an elementary charge. The expression for $\epsilon$ is:
\begin{equation}
\epsilon{(\mathbf{r}, \mathbf{r}'; \omega)} = \delta{(\mathbf{r} - \mathbf{r}')} - 
\int d\mathbf{r}'' v{(\mathbf{r} - \mathbf{r}'')} P{(\mathbf{r}'', \mathbf{r}'; \omega)}
\label{eqn:eps}
\end{equation}
where $\delta$ is the Dirac delta function and $P$ is the polarizability. 
The latter is evaluated within the random phase approximation (RPA):
\begin{equation}
P{(\mathbf{r}, \mathbf{r}'; \omega)} = - \frac{i}{2\pi} \int d\omega' 
G{(\mathbf{r}, \mathbf{r}'; \omega + \omega')} G{(\mathbf{r}', \mathbf{r}; \omega')}
\label{eqn:p}
\end{equation}
Calculations are performed in reciprocal space, for instance 
$\epsilon{(\mathbf{r}, \mathbf{r}'; \omega)}$ is Fourier transformed 
to $\epsilon_{\mathbf{GG}'}{(\mathbf{q}; \omega)}$, where $\mathbf{G}$ 
is the reciprocal lattice vector and $\mathbf{q}$ is the Bloch 
wave vector.

In practice, the $GW$ method is applied perturbatively on top 
of Kohn-Sham density functional theory (DFT) \cite{kohn-65-dft} 
calculations. It is often assumed that the Kohn-Sham orbitals 
$\psi_{n\mathbf{k}}^\mathrm{KS}{(\mathbf{r})}$ are good approximation for 
the quasiparticle orbitals $\psi_{n\mathbf{k}}^\mathrm{QP}{(\mathbf{r})}$. 
$\Sigma$ is then diagonal in the basis of $\psi_{n\mathbf{k}}^\mathrm{KS}$ 
and the quasiparticle energies are expressed by \cite{hybertsen-86-gw}:
\begin{eqnarray}
E_{n\mathbf{k}}^\mathrm{QP} = E_{n\mathbf{k}}^\mathrm{KS} &+& 
\left< \psi_{n\mathbf{k}}^\mathrm{KS}{(\mathbf{r})} \right| 
\Sigma{(\mathbf{r}, \mathbf{r}'; E_{n\mathbf{k}}^\mathrm{QP})} \nonumber \\
&-& V_\mathrm{xc}{[\rho_\mathrm{scf}{(\mathbf{r})}]}{(\mathbf{r})} 
\delta{(\mathbf{r} - \mathbf{r}')} 
\left| \psi_{n\mathbf{k}}^\mathrm{KS}{(\mathbf{r}')} \right>
\label{eqn:eqp}
\end{eqnarray}
where $E_{n\mathbf{k}}^\mathrm{KS}$ are the Kohn-Sham energies, 
$V_\mathrm{xc}$ is the exchange-correlation potential, and 
$\rho_\mathrm{scf}$ is the self-consistent charge density.

The Kohn-Sham ansatz is often used in conjunction with 
\textit{ab initio} pseudopotentials \cite{hamann-79-pp} 
assuming separation of electrons into core and valence states. 
This implies that the $\Sigma$ and $V_\mathrm{xc}$ terms 
of Eq.~(\ref{eqn:eqp}) only include contributions from 
the valence states, while contributions from the core states are treated 
at the DFT level in the $E_{n\mathbf{k}}^\mathrm{KS}$ term 
of Eq.~(\ref{eqn:eqp}), and the core-valence interaction 
is neglected \cite{hedin-70-gw, hybertsen-86-gw}. 
The latter is of particular concern 
when core and valence orbitals overlap, 
such as would occur in Zn if 
1s$^2$2s$^2$2p$^6$3s$^2$3p$^6$ states 
were treated as core states and 
3d$^{10}$4s$^2$ states as valence states. 
The core-valence interaction can be included at the DFT level 
using the non-linear core correction (NLCC) \cite{louie-82-nlcc} 
which introduces the partial core charge density $\rho_\mathrm{core}$ 
in the evaluation of the exchange-correlation potential, 
$V_\mathrm{xc}{[\rho_\mathrm{core}+\rho_\mathrm{scf}]}$. 
The $GW$ method on the other hand requires the entire shell 
of semicore states (such as 3s$^2$3p$^6$3d$^{10}$ states in Zn) 
to be explicitly treated as valence states in order 
to eliminate errors due to neglecting the core-valence 
interaction \cite{rohlfing-95-semicore, rohlfing-97-semicore, 
marini-01-semicore, kang-10-tio2-gw, umari-12-semicore}. 
All calculations in this work are performed treating the 
entire third shells of Zn, Cu, and Ti as valence states.

The core-valence partitioning brings up another issue, 
namely that the charge density used for the evaluation 
of the $V_\mathrm{xc}$ term in Eq.~(\ref{eqn:eqp}) 
must be consistent with the orbitals used in the 
construction of the $\Sigma$ operator in the said 
equation \cite{arnaud-00-gw-val, fleszar-05-gw-nlcc, 
gomez-abal-08-gw-ae, li-12-gw-ae}. 
In particular, it was shown that if the NLCC is used 
in the DFT calculation, $\rho_\mathrm{core}$ must be 
set to zero when evaluating the $V_\mathrm{xc}$ term 
of Eq.~(\ref{eqn:eqp}) \cite{fleszar-05-gw-nlcc}. 
To study the effect of imbalance between the $\Sigma$ 
and $V_\mathrm{xc}$ terms in Eq.~(\ref{eqn:eqp}), 
we use $\rho_\mathrm{core}$ derived from the deep 
core states (such as 2s$^2$2p$^6$ states in Zn). 
Even though there is negligible overlap 
between the deep core and semicore orbitals 
(such as the second and third shells of Zn), 
the integrated partial core charge 
$q_\mathrm{core} = \int d\mathbf{r} 
\rho_\mathrm{core}{(\mathbf{r})}$ 
is not small ($q_\mathrm{core} = 7.67 e$ in Zn). 
In what follows we examine how keeping $\rho_\mathrm{core}$ 
in the $V_\mathrm{xc}$ term of Eq.~(\ref{eqn:eqp}) 
affects the results of $GW$ calculations as compared 
to the case of zeroing out $\rho_\mathrm{core}$ 
in the $V_\mathrm{xc}$ term.

Several different approaches have been developed for constructing 
$\Sigma$ and calculating its matrix elements entering Eq.~(\ref{eqn:eqp}):
\begin{itemize}
\item Non-self-consistent $G_0W_0$ scheme \cite{hybertsen-86-gw} 
when $G$ and $P$ are obtained by plugging $\psi_{n\mathbf{k}}^\mathrm{KS}$ 
and $E_{n\mathbf{k}}^\mathrm{KS}$ into Eqs.~(\ref{eqn:g}) and~(\ref{eqn:p}).
\item Eigenvalue self-consistent $GW$ scheme \cite{shishkin-07-zno-gw} 
when $G$ and $P$ are constructed from $\psi_{n\mathbf{k}}^\mathrm{KS}$ 
and $E_{n\mathbf{k}}^\mathrm{QP}$, the latter being determined 
iteratively starting from $E_{n\mathbf{k}}^\mathrm{KS}$.
\item Eigenvalue self-consistent $GW_0$ scheme \cite{shishkin-07-zno-gw} 
when $G$ is calculated using $\psi_{n\mathbf{k}}^\mathrm{KS}$ 
and $E_{n\mathbf{k}}^\mathrm{QP}$ while $P$ is calculated using 
$\psi_{n\mathbf{k}}^\mathrm{KS}$ and $E_{n\mathbf{k}}^\mathrm{KS}$.
\item Eigenvector self-consistent $GW$ schemes 
\cite{faleev-04-scgw, van-schilfgaarde-06-scgw, bruneval-06-scgw} 
when $\psi_{n\mathbf{k}}^\mathrm{QP}$ are constructed iteratively 
using off-diagonal matrix elements of $\Sigma$ in the basis of 
$\psi_{n\mathbf{k}}^\mathrm{KS}$.
\end{itemize}
It was shown that the self-consistent $GW$ scheme without 
the vertex correction in $\Sigma$ (beyond the $GW$ approximation) 
overestimates the experimental band gaps \cite{schone-98-scgw}. 
Better agreement with experiment is obtained using the 
$GW_0$ scheme because the effects of self-consistency 
in $W$ and of vertex correction in $\Sigma$ largely 
cancel out \cite{shirley-96-gw-vertex, 
von-barth-96-gw-vertex, holm-98-gw-vertex}. 
It should be noted that the self-consistency in $G$ 
without the vertex correction in $\Sigma$ violates 
the Ward-Takahashi identity representing 
the local electron number conservation law 
\cite{takada-01-gw-vertex}. 
For the purpose of this work, 
we employ non-self-consistent $G_0W_0$ 
and eigenvalue self-consistent $GW_0$ schemes.

The energy integral in Eq.~(\ref{eqn:sigma}) can be evaluated 
by direct numerical integration \cite{deslippe-12-bgw}, 
employing the Hilbert transform \cite{shishkin-06-vasp-gw}, 
the contour deformation technique \cite{lebegue-03-cd}, 
or using a plasmon-pole model (PPM) to approximate 
the $\omega$ dependence of $\epsilon^{-1}$. 
The first three methods are thereafter referred as non-PPM. 
Two popular choices for PPM are 
the Hybertsen-Louie (HL) PPM \cite{hybertsen-86-gw, zhang-89-gpp-noinv} 
and the Godby-Needs (GN) PPM \cite{godby-89-ppm}. 
The HL PPM takes as input the static inverse dielectric function 
$\epsilon^{-1}$ at $\omega = 0$ and the charge density $\rho_\mathrm{ppm}$ 
which is used to compute the effective bare plasma frequencies. 
The GN PPM takes as input $\epsilon^{-1}$ at two frequencies, 
$\omega = 0$ and $\omega = i\Omega$, where $\Omega$ is a parameter. 
The HL PPM recently came under criticism for poorly reproducing the 
$\omega$ dependence of the RPA $\epsilon^{-1}$ as compared to the 
GN PPM \cite{stankovski-11-zno-gw, miglio-12-zno-gw, larson-13-zno-ppm}.

As we show in this paper, the poor performance of the HL PPM 
stems from the improper choice of $\rho_\mathrm{ppm}$. 
One sensible choice for $\rho_\mathrm{ppm}$ 
is the charge density of the valence electrons 
(oxygen 2p$^6$ states), $\rho_\mathrm{ppm}=\rho_\mathrm{val}$, 
owing to the fact that the dielectric screening is dominated 
by the valence electrons \cite{kaur-13-gw-dielec}. 
This choice for $\rho_\mathrm{ppm}$ was implicitly assumed 
in the original derivation of the HL PPM \cite{hybertsen-86-gw}. 
Another common choice for $\rho_\mathrm{ppm}$ 
is the self-consistent charge density, 
$\rho_\mathrm{ppm}=\rho_\mathrm{scf}$, which includes the core 
electrons treated as valence in the construction of the pseudopotentials 
(oxygen 2s$^2$ states and the transition metal third shell). 
Our calculations demonstrate that the HL PPM approaches the GN PPM and 
the RPA results when $\rho_\mathrm{ppm}$ is set to $\rho_\mathrm{val}$. 
At the same time, the poor performance of the HL PPM discussed in the 
literature \cite{stankovski-11-zno-gw, miglio-12-zno-gw, larson-13-zno-ppm} 
is attributed to setting $\rho_\mathrm{ppm}$ equal to $\rho_\mathrm{scf}$.

\begin{table}
\caption{Pseudopotential parameters for Zn$^{2+}$, Cu$^{2+}$, Ti$^{2+}$, and O. 
Shown are the electronic core and valence configurations, the integrated partial 
core charge $q_\mathrm{core} = \int d\mathbf{r} \rho_\mathrm{core}{(\mathbf{r})}$, 
the partial core radius $r_\mathrm{core}$ determined by the condition 
$\rho_\mathrm{core}{(r_\mathrm{core})} = 2 \rho_\mathrm{val}{(r_\mathrm{core})}$, 
and the matching radii for different angular momentum channels $r_\mathrm{s,p,d}$. 
Core charge is in units of elementary charge, all radii are in Bohr.
\label{tab:pp}}
\begin{ruledtabular}
\begin{tabular}{lllccccc}
~         & Core               & Valence               & $q_\mathrm{core}$ & $r_\mathrm{core}$ & $r_\mathrm{s}$ & $r_\mathrm{p}$ & $r_\mathrm{d}$ \\
\hline
Zn$^{2+}$ & 1s$^2$2s$^2$2p$^6$ & 3s$^2$3p$^6$3d$^{10}$ & 7.67              & 0.31              & 1.00           &         1.00   & 0.85           \\
Cu$^{2+}$ & 1s$^2$2s$^2$2p$^6$ & 3s$^2$3p$^6$3d$^9$    & 7.53              & 0.33              & 1.05           &         1.05   & 0.90           \\
Ti$^{2+}$ & 1s$^2$2s$^2$2p$^6$ & 3s$^2$3p$^6$3d$^2$    & 6.27              & 0.52              & 1.20           &         1.25   & 1.35           \\
O         & 1s$^2$             & 2s$^2$2p$^4$          & 1.37              & 0.34              & 1.10           &         1.10   &                \\
\end{tabular}
\end{ruledtabular}
\end{table}

\begin{table}
\caption{Parameters of DFT and $GW$ calculations for wurtzite ZnO, cuprite Cu$_2$O, 
and rutile TiO$_2$. MP stands for a Monkhorst-Pack grid \cite{monkhorst-76-kgrid} 
for summing over the Brillouin zone to obtain $\rho_\mathrm{scf}$, $\rho_\mathrm{val}$, 
$\epsilon$, and $\Sigma$. $E_\psi$, $E_v$, $E_\epsilon$, and $E_W$ are kinetic energy 
cutoffs for the plane wave expansion of $\psi_{n\mathbf{k}}^\mathrm{KS}$, $v$, $\epsilon$, 
and $W$, respectively. $N_\mathrm{KS}$ is the number of Kohn-Sham bands (both occupied 
and unoccupied) with the energies up to about $E_\mathrm{KS}$ above the average 
($\mathbf{G} = \mathbf{0}$ component) electrostatic (ionic plus Hartree) potential.
\label{tab:param}}
\begin{ruledtabular}
\begin{tabular}{lccc}
~                          & Wurtzite ZnO        & Cuprite Cu$_2$O     & Rutile TiO$_2$      \\
\hline
MP $\rho_\mathrm{scf,val}$ & 9$\times$9$\times$5 & 7$\times$7$\times$7 & 6$\times$6$\times$9 \\
MP $\epsilon,\Sigma$       & 5$\times$5$\times$3 & 4$\times$4$\times$4 & 3$\times$3$\times$5 \\
$E_{\psi,v}$ (Ry)          & 400                 & 350                 & 250                 \\
$E_{\epsilon,W}$ (Ry)      & 80                  & 80                  & 80                  \\
$N_\mathrm{KS}$            & 1500                & 2400                & 1900                \\
$E_\mathrm{KS}$ (Ry)       & 40                  & 40                  & 40                  \\
\end{tabular}
\end{ruledtabular}
\end{table}

\begin{table}
\caption{Structural parameters of wurtzite ZnO, cuprite Cu$_2$O, and rutile TiO$_2$ 
measured by X-ray diffraction \cite{kihara-85-zno-struc, kirfel-90-cu2o-struc, 
abrahams-71-tio2-struc} and calculated using DFT with LDA and GGA exchange-correlation 
functionals.
\label{tab:struc}}
\begin{ruledtabular}
\begin{tabular}{lccccccc}
~     & \multicolumn{3}{c}{Wurtzite ZnO} & Cuprite Cu$_2$O & \multicolumn{3}{c}{Rutile TiO$_2$} \\
~     & $a$ (\AA) & $c$ (\AA) & $u$      & $a$ (\AA)       & $a$ (\AA) & $c$ (\AA) & $u$        \\
\hline
X-ray\footnote{From Refs.~\onlinecite{kihara-85-zno-struc, kirfel-90-cu2o-struc, abrahams-71-tio2-struc}.}
      & 3.25      & 5.20      & 0.382    & 4.27            & 4.59      & 2.96      & 0.305      \\
LDA   & 3.19      & 5.16      & 0.378    & 4.18            & 4.56      & 2.92      & 0.304      \\
GGA   & 3.28      & 5.30      & 0.379    & 4.31            & 4.65      & 2.97      & 0.305      \\
\end{tabular}
\end{ruledtabular}
\end{table}

\section{Computational details}
\label{sec:comp}

\begin{figure}
\includegraphics[width=202.298pt]{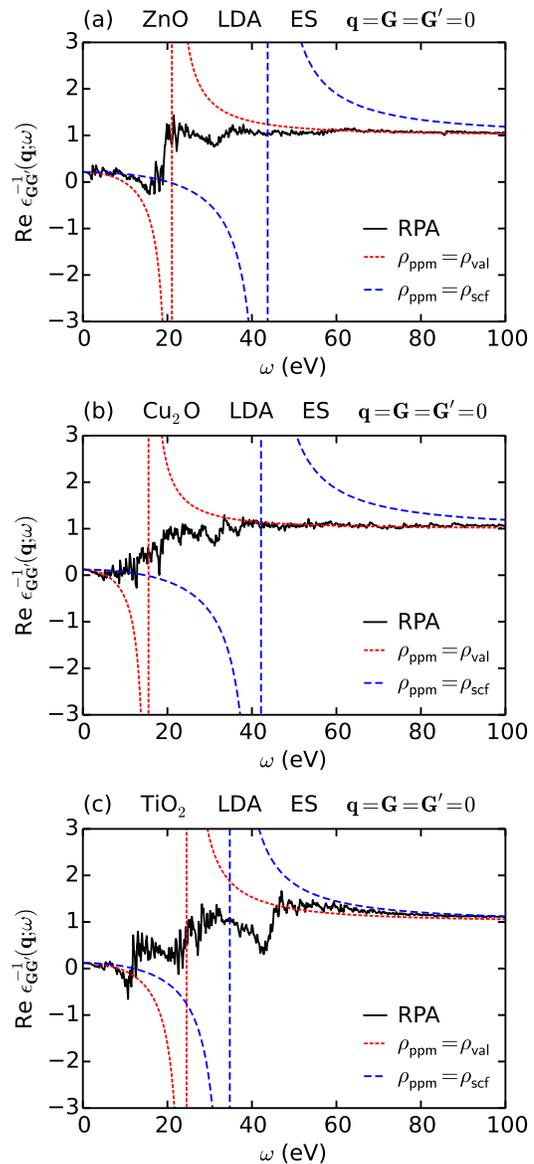}
\caption{Real parts of inverse dielectric functions $\epsilon_{{\bf GG}'}^{-1}{({\bf q}; \omega)}$ 
at ${\bf q} = {\bf G} = {\bf G}' = {\bf 0}$ of (a) wurtzite ZnO, (b) cuprite Cu$_2$O, 
and (c) rutile TiO$_2$ in case of the LDA starting point and experimental structural 
parameters (ES) calculated within the RPA (solid black) and constructed using the HL PPM with 
$\rho_\mathrm{ppm}=\rho_\mathrm{val}$ (short-dashed red) and $\rho_\mathrm{ppm}=\rho_\mathrm{scf}$ 
(long-dashed blue). The HL PPM mode frequencies $\tilde{\omega}_{{\bf GG}'}{({\bf q})}$ are 
shown by the vertical dashed lines at (a) 21.0 eV and 43.8 eV, (b) 15.6 eV and 42.1 eV, 
and (c) 24.6 eV and 34.7 eV.
\label{fig:epsinv}}
\end{figure}

\begin{figure}
\includegraphics[width=246.000pt]{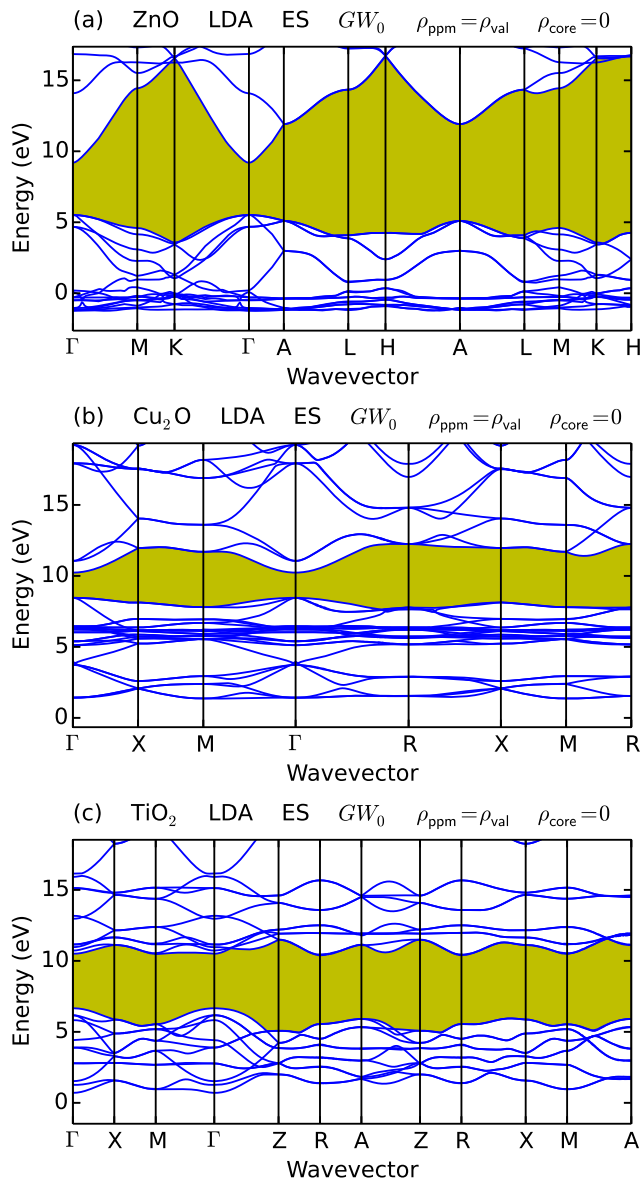}
\caption{Quasiparticle band structures of (a) wurtzite ZnO, (b) cuprite Cu$_2$O, and (c) rutile TiO$_2$ 
calculated using the LDA starting point, experimental structural parameters (ES), the eigenvalue 
self-consistent $GW_0$ scheme, the HL PPM with $\rho_\mathrm{ppm}=\rho_\mathrm{val}$, and matrix 
elements of $V_\mathrm{xc}$ without NLCC ($\rho_\mathrm{core} = 0$). The zero reference for the energy 
scale is the average (${\bf G} = {\bf 0}$ component) electrostatic (ionic plus Hartree) potential. 
The ${\bf k}$-point labeling is from Ref.~\onlinecite{setyawan-10-kpoint}. The band gaps are 
shaded in yellow.
\label{fig:band}}
\end{figure}

To examine the effects of different approximations discussed in 
Sec.~\ref{sec:theor} on the quasiparticle band gaps and band edges 
of TMOs, we perform a series of calculations for wurtzite ZnO, 
cuprite Cu$_2$O, and rutile TiO$_2$ using \texttt{Quantum ESPRESSO} 
\cite{giannozzi-09-qe} and \texttt{BerkeleyGW} \cite{deslippe-12-bgw} 
codes for the DFT and $GW$ parts, respectively. Calculations are carried 
out for the spin-unpolarized case with the local density approximation (LDA) 
in the PW form \cite{perdew-92-pw} and the generalized gradient approximation 
(GGA) in the PBE form \cite{perdew-96-pbe} for the exchange-correlation 
functional. Norm-conserving pseudopotentials are generated in a separable 
non-local form \cite{kleinman-82-pp} using the RRKJ scheme \cite{rappe-90-pp} 
and including scalar relativistic corrections and non-linear core corrections 
(NLCC) \cite{louie-82-nlcc}. The pseudopotential parameters are summarized 
in Table~\ref{tab:pp}. Convergence studies with respect to the size of 
the Monkhorst-Pack grid \cite{monkhorst-76-kgrid}, kinetic energy cutoffs, 
and the number of unoccupied Kohn-Sham bands used in the calculation of 
$\epsilon$ and $\Sigma$ are reported elsewhere \cite{shih-10-zno-gw, 
friedrich-11-zno-gw, stankovski-11-zno-gw, deslippe-13-sr, 
malashevich-14-tio2-gw}. The parameters used in our 
calculations are summarized in Table~\ref{tab:param}. 
The Monkhorst-Pack grids for $\rho_\mathrm{scf}$, $\rho_\mathrm{val}$, and 
$\Sigma$ are $\Gamma$-centered and the ones for $\epsilon$ are shifted 
by half a grid spacing in all directions. A small wave vector along the 
(111) direction in crystal coordinates is used to calculate $\epsilon$ 
at the $\Gamma$ point. The convergence of $\Sigma$ with respect 
to the size of the Monkhorst-Pack grid is accelerated 
by averaging $v$ and $W$ inside the Voronoi cells 
of the $\left(\mathbf{k}+\mathbf{G}\right)$-points 
near the $\Gamma$-point \cite{deslippe-12-bgw}. 
The convergence of $\Sigma$ with respect to the number 
of unoccupied Kohn-Sham bands is accelerated by using 
the static remainder correction \cite{deslippe-13-sr}. 
To ensure convergence of the stress tensor, 
structural relaxations are performed using 3 times higher 
kinetic energy cutoffs than those listed in Table~\ref{tab:param}. 
The experimental and theoretical structural parameters 
(thereafter referred to as ES and TS, respectively) 
are listed in Table~\ref{tab:struc}.

Special consideration is required when 
constructing $\rho_\mathrm{val}$ used in the HL PPM. 
Given the two formula units per primitive cell and the electronic valence 
configurations listed in Table~\ref{tab:pp}, ZnO, Cu$_2$O, and TiO$_2$ 
have 26, 44, and 24 valence bands, respectively. The top of the valence 
manifold is derived from the oxygen 2p$^6$ states: bands 21--26 in ZnO, 
bands 39--44 in Cu$_2$O, and bands 13--24 in TiO$_2$. The lower valence 
bands are derived from the oxygen 2s$^2$ states and the transition metal 
third shell: bands 1--20 from O 2s$^2$ \& Zn 3s$^2$3p$^6$3d$^{10}$ 
in ZnO, bands 1--38 from O 2s$^2$ \& Cu 3s$^2$3p$^6$3d$^{10}$ in Cu$_2$O, 
and bands 1--12 from O 2s$^2$ \& Ti 3s$^2$3p$^6$ in TiO$_2$. 
In TiO$_2$ the oxygen 2p states are separated from the transition metal 
3d states by an energy gap, while in ZnO and Cu$_2$O they overlap. 
These overlapping states should be decoupled in order to unambiguously 
construct $\rho_\mathrm{val}$ from the oxygen 2p$^6$ states. 
For that purpose we employ the DFT+U method with the following parameters: 
$U = 8.0$ eV and $J = 0.9$ eV for ZnO \cite{shih-10-zno-gw}; 
$U = 7.5$ eV and $J = 0.98$ eV for Cu$_2$O \cite{heinemann-13-cu2o-hse}. 
Note that the DFT+U method is only used for constructing $\rho_\mathrm{val}$, 
while $GW$ calculations are carried out starting from DFT orbitals. 
To quantify the effect of $U$, we perform two sets of $GW$ calculations, 
one using DFT $\rho_\mathrm{val}$ and another using DFT+U $\rho_\mathrm{val}$. 
It is found that the inclusion of $U$ in $\rho_\mathrm{val}$ only 
changes the $GW$ band gaps by 10 meV and the $GW$ band edges by 40 meV. 
The much larger effect of using $\rho_\mathrm{scf}$ in the HL PPM 
will be discussed in Sec.~\ref{sec:res}.

Let us now describe the implementation of the 
eigenvalue self-consistent $GW_0$ scheme. Iterations on 
$E_{n\mathbf{k}}^\mathrm{QP}$ entering Eq.~(\ref{eqn:g}) 
are performed by explicitly calculating the matrix elements 
of $\Sigma$ and the values of $E_{n\mathbf{k}}^\mathrm{QP}$ 
for several valence and conduction bands near the band edges 
(16 valence and 14 conduction for ZnO, 26 valence and 10 
conduction for Cu$_2$O, 12 valence and 16 conduction for TiO$_2$) 
and by applying the $\mathbf{k}$-dependent scissors operators 
to the lower valence and higher conduction bands. 
The $\mathbf{k}$-dependent scissor shifts are obtained from 
the lowest valence and highest conduction bands for which 
the matrix elements of $\Sigma$ and the values of 
$E_{n\mathbf{k}}^\mathrm{QP}$ are explicitly calculated. 
It is found that performing four iterations is sufficient 
to converge $E_{n\mathbf{k}}^\mathrm{QP}$ to within 10 meV.

\section{Results and discussion}
\label{sec:res}

\begin{figure*}
\includegraphics[width=431.381pt]{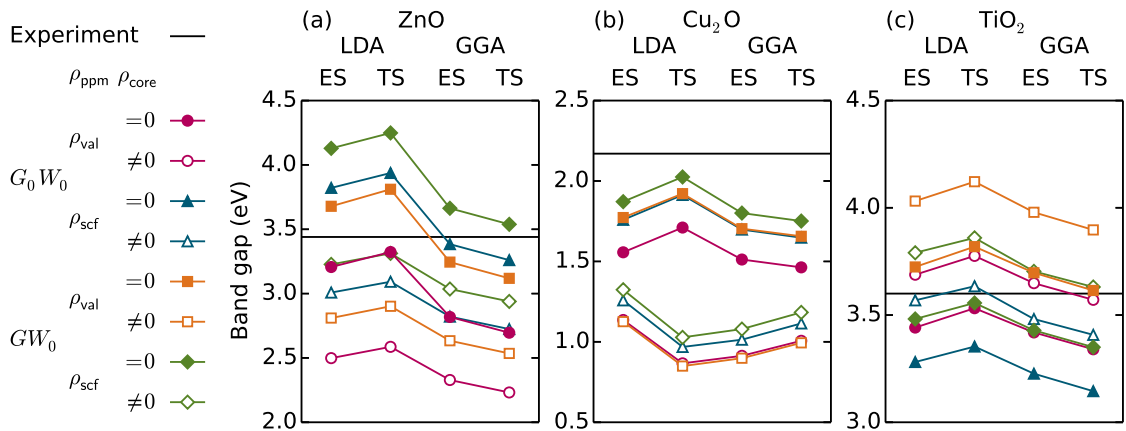}
\caption{Quasiparticle band gaps of (a) wurtzite ZnO, 
(b) cuprite Cu$_2$O, and (c) rutile TiO$_2$ calculated 
within the $GW$ method and plotted as a function of the 
starting point (obtained from the LDA or GGA calculations) 
and of the structural parameters (either experimental 
or theoretical, labeled as ES and TS, respectively). 
Different symbols indicate the values calculated 
using different flavors of the $GW$ method, as shown 
in the legend on the left. The experimental band gaps 
taken from Refs.~\onlinecite{reynolds-99-zno-bandgap, 
baumeister-61-cu2o-bandgap, rangan-10-tio2-bandgap} 
are shown by horizontal lines.
\label{fig:gapvis}}
\end{figure*}

\begin{table}
\caption{Band gaps of wurtzite ZnO measured using photoluminescence (PL) \cite{reynolds-99-zno-bandgap} 
and calculated using DFT and $GW$. DFT and $GW$ band gaps are obtained using different 
exchange-correlation functionals (LDA and GGA), experimental and theoretical structural 
parameters (ES and TS), non-self-consistent $G_0W_0$ and eigenvalue self-consistent $GW_0$ 
schemes, HL PPM with $\rho_\mathrm{ppm} = \rho_\mathrm{val}$ and $\rho_\mathrm{scf}$, and 
matrix elements of $V_\mathrm{xc}$ without and with NLCC ($\rho_\mathrm{core} = 0$ and $\neq 0$). 
The DFT and $GW$ band gaps are direct at the $\Gamma$ point. All values are in eV.
\label{tab:znogap}}
\begin{ruledtabular}
\begin{tabular}{lllcccc}
\multicolumn{3}{l}{Wurtzite ZnO}                                                        & \multicolumn{2}{c}{LDA}       & \multicolumn{2}{c}{GGA}       \\
~                         & $\rho_\mathrm{ppm}$                  & $\rho_\mathrm{core}$ & ES            & TS            & ES            & TS            \\
\hline
\multicolumn{3}{l}{PL\footnote{From Ref.~\onlinecite{reynolds-99-zno-bandgap}.}}        & \multicolumn{4}{c}{3.44}                                      \\
\multicolumn{3}{l}{DFT}                                                                 &         0.74  &         0.80  &         0.85  &         0.78  \\
\multirow{4}{*}{$G_0W_0$} & \multirow{2}{*}{$\rho_\mathrm{val}$} & $=0$                 &         3.21  &         3.32  &         2.82  &         2.70  \\
                          &                                      & $\neq0$              &         2.50  &         2.59  &         2.33  &         2.23  \\
                          & \multirow{2}{*}{$\rho_\mathrm{scf}$} & $=0$                 &         3.82  &         3.94  &         3.38  &         3.26  \\
                          &                                      & $\neq0$              &         3.01  &         3.09  &         2.82  &         2.72  \\
\multirow{4}{*}{$GW_0$}   & \multirow{2}{*}{$\rho_\mathrm{val}$} & $=0$                 &         3.68  &         3.81  &         3.24  &         3.12  \\
                          &                                      & $\neq0$              &         2.81  &         2.90  &         2.63  &         2.54  \\
                          & \multirow{2}{*}{$\rho_\mathrm{scf}$} & $=0$                 &         4.13  &         4.25  &         3.66  &         3.54  \\
                          &                                      & $\neq0$              &         3.23  &         3.31  &         3.04  &         2.94  \\
\end{tabular}
\end{ruledtabular}
\end{table}

\begin{table}
\caption{Band gaps of cuprite Cu$_2$O measured using optical absorption spectroscopy (OAS) 
\cite{baumeister-61-cu2o-bandgap} and calculated using DFT and $GW$. DFT and $GW$ band 
gaps are obtained using different exchange-correlation functionals (LDA and GGA), 
experimental and theoretical structural parameters (ES and TS), non-self-consistent 
$G_0W_0$ and eigenvalue self-consistent $GW_0$ schemes, HL PPM with 
$\rho_\mathrm{ppm} = \rho_\mathrm{val}$ and $\rho_\mathrm{scf}$, and matrix elements 
of $V_\mathrm{xc}$ without and with NLCC ($\rho_\mathrm{core} = 0$ and $\neq 0$). The 
DFT and $GW$ band gaps are direct at the $\Gamma$ point. All values are in eV.
\label{tab:cuogap}}
\begin{ruledtabular}
\begin{tabular}{lllcccc}
\multicolumn{3}{l}{Cuprite Cu$_2$O}                                                     & \multicolumn{2}{c}{LDA}       & \multicolumn{2}{c}{GGA}       \\
~                         & $\rho_\mathrm{ppm}$                  & $\rho_\mathrm{core}$ & ES            & TS            & ES            & TS            \\
\hline
\multicolumn{3}{l}{OAS\footnote{From Ref.~\onlinecite{baumeister-61-cu2o-bandgap}.}}    & \multicolumn{4}{c}{2.17}                                      \\
\multicolumn{3}{l}{DFT}                                                                 &         0.52  &         0.69  &         0.53  &         0.47  \\
\multirow{4}{*}{$G_0W_0$} & \multirow{2}{*}{$\rho_\mathrm{val}$} & $=0$                 &         1.56  &         1.71  &         1.51  &         1.46  \\
                          &                                      & $\neq0$              &         1.14  &         0.87  &         0.91  &         1.01  \\
                          & \multirow{2}{*}{$\rho_\mathrm{scf}$} & $=0$                 &         1.76  &         1.91  &         1.70  &         1.65  \\
                          &                                      & $\neq0$              &         1.26  &         0.97  &         1.01  &         1.11  \\
\multirow{4}{*}{$GW_0$}   & \multirow{2}{*}{$\rho_\mathrm{val}$} & $=0$                 &         1.77  &         1.92  &         1.70  &         1.66  \\
                          &                                      & $\neq0$              &         1.13  &         0.85  &         0.90  &         0.99  \\
                          & \multirow{2}{*}{$\rho_\mathrm{scf}$} & $=0$                 &         1.87  &         2.03  &         1.80  &         1.75  \\
                          &                                      & $\neq0$              &         1.32  &         1.03  &         1.08  &         1.18  \\
\end{tabular}
\end{ruledtabular}
\end{table}

\begin{table}
\caption{Band gaps of rutile TiO$_2$ measured using photoemission spectroscopy (PES) 
\cite{rangan-10-tio2-bandgap} and calculated using DFT and $GW$. DFT and $GW$ band 
gaps are obtained using different exchange-correlation functionals (LDA and GGA), 
experimental and theoretical structural parameters (ES and TS), non-self-consistent 
$G_0W_0$ and eigenvalue self-consistent $GW_0$ schemes, HL PPM with 
$\rho_\mathrm{ppm} = \rho_\mathrm{val}$ and $\rho_\mathrm{scf}$, and matrix elements 
of $V_\mathrm{xc}$ without and with NLCC ($\rho_\mathrm{core} = 0$ and $\neq 0$). 
The DFT and $GW$ band gaps are direct at the $\Gamma$ point (in regular font) 
and indirect between the $\Gamma$ point at the VBM and the R point at the CBM 
(in cursive font). All values are in eV.
\label{tab:tiogap}}
\begin{ruledtabular}
\begin{tabular}{lllcccc}
\multicolumn{3}{l}{Rutile TiO$_2$}                                                      & \multicolumn{2}{c}{LDA}       & \multicolumn{2}{c}{GGA}       \\
~                         & $\rho_\mathrm{ppm}$                  & $\rho_\mathrm{core}$ & ES            & TS            & ES            & TS            \\
\hline
\multicolumn{3}{l}{PES\footnote{From Ref.~\onlinecite{rangan-10-tio2-bandgap}.}}        & \multicolumn{4}{c}{3.60}                                      \\
\multicolumn{3}{l}{DFT}                                                                 &         1.82  &         1.85  &         1.90  &         1.85  \\
\multirow{4}{*}{$G_0W_0$} & \multirow{2}{*}{$\rho_\mathrm{val}$} & $=0$                 & \textit{3.44} & \textit{3.53} & \textit{3.42} & \textit{3.34} \\
                          &                                      & $\neq0$              & \textit{3.69} & \textit{3.78} & \textit{3.65} & \textit{3.57} \\
                          & \multirow{2}{*}{$\rho_\mathrm{scf}$} & $=0$                 & \textit{3.28} & \textit{3.35} & \textit{3.23} &         3.14  \\
                          &                                      & $\neq0$              & \textit{3.57} & \textit{3.63} & \textit{3.48} &         3.41  \\
\multirow{4}{*}{$GW_0$}   & \multirow{2}{*}{$\rho_\mathrm{val}$} & $=0$                 & \textit{3.72} & \textit{3.82} & \textit{3.70} & \textit{3.61} \\
                          &                                      & $\neq0$              & \textit{4.03} & \textit{4.12} & \textit{3.98} & \textit{3.90} \\
                          & \multirow{2}{*}{$\rho_\mathrm{scf}$} & $=0$                 & \textit{3.48} & \textit{3.56} & \textit{3.43} &         3.35  \\
                          &                                      & $\neq0$              & \textit{3.79} & \textit{3.86} & \textit{3.70} & \textit{3.63} \\
\end{tabular}
\end{ruledtabular}
\end{table}

$GW$ calculations for wurtzite ZnO, cuprite Cu$_2$O, 
and rutile TiO$_2$ are performed using LDA and 
GGA starting points, experimental and theoretical 
structural parameters (ES and TS), non-self-consistent 
$G_0W_0$ and eigenvalue self-consistent $GW_0$ schemes, 
HL PPM with $\rho_\mathrm{ppm}$ set to DFT+U $\rho_\mathrm{val}$ 
and DFT $\rho_\mathrm{scf}$, and matrix elements of $V_\mathrm{xc}$ 
without and with NLCC ($\rho_\mathrm{core} = 0$ and $\neq 0$). 
In the latter case, values of integrated partial core charge 
$q_\mathrm{core} = \int d\mathbf{r} \rho_\mathrm{core}{(\mathbf{r})}$ 
are listed in Table~\ref{tab:pp}. 
Fig.~\ref{fig:epsinv} shows the real parts of 
$\epsilon_\mathbf{00}^{-1}{(\mathbf{0}; \omega)}$ 
for the three TMOs in case of the LDA starting point 
and experimental structural parameters (ES) calculated 
within the RPA and constructed using the HL PPM with 
$\rho_\mathrm{ppm}=\rho_\mathrm{val}$ and $\rho_\mathrm{scf}$. 
Fig.~\ref{fig:band} shows the quasiparticle band structures 
calculated using the LDA starting point, experimental structural 
parameters (ES), the eigenvalue self-consistent $GW_0$ scheme, 
the HL PPM with $\rho_\mathrm{ppm}=\rho_\mathrm{val}$, and matrix 
elements of $V_\mathrm{xc}$ without NLCC ($\rho_\mathrm{core}=0$).
Fig.~\ref{fig:gapvis} shows the quasiparticle band gaps plotted 
as a function of the starting point (obtained from the LDA or GGA 
calculations) and of the structural parameters (either experimental 
or theoretical, labeled as ES and TS, respectively). Different 
symbols indicate the values calculated using different flavors 
of the $GW$ method. In this context, flavor refers to the 
choice of self-consistent scheme, $\rho_\mathrm{ppm}$, and 
$\rho_\mathrm{core}$. The experimental band gaps taken from 
Refs.~\onlinecite{reynolds-99-zno-bandgap, baumeister-61-cu2o-bandgap, 
rangan-10-tio2-bandgap} are shown for comparison. 
Tables~\ref{tab:znogap}-\ref{tab:tiogap} give the experimental 
and calculated band gaps plotted in Fig.~\ref{fig:gapvis} 
as well as the Kohn-Sham values not shown in Fig.~\ref{fig:gapvis}. 
Kohn-Sham and quasiparticle band energies $E_{n\mathbf{k}}^\mathrm{KS}$ 
and $E_{n\mathbf{k}}^\mathrm{QP}$ and matrix elements of $V_\mathrm{xc}$ 
and $\Sigma$ at the valence band maximum (VBM) and conduction band 
minimum (CBM) are provided in Supplemental Material \cite{nlcc_suppl}.

Comparing Fig.~\ref{fig:epsinv}(a) with 
Fig.~2(a) of Ref.~\onlinecite{stankovski-11-zno-gw}, 
Fig.~5(a) of Ref.~\onlinecite{miglio-12-zno-gw}, 
and Fig.~4(d) of Ref.~\onlinecite{larson-13-zno-ppm}, 
we find that in the case of ZnO, 
the HL PPM becomes similar to the GN PPM when 
$\rho_\mathrm{ppm}$ is set to $\rho_\mathrm{val}$. 
One can see from Fig.~\ref{fig:epsinv} that 
for all three oxides, 
the HL PPM with $\rho_\mathrm{ppm}=\rho_\mathrm{val}$ 
gives a better fit to the RPA results than 
the HL PPM with $\rho_\mathrm{ppm}=\rho_\mathrm{scf}$. 
We note that the HL PPM suffers from the ambiguity 
of constructing the proper $\rho_\mathrm{ppm}$. 
This problem is absent in the GN PPM, 
suggesting that the HL PPM is more difficult 
to use for studying TMOs than the GN PPM.

Several conclusions can be drawn from statistical 
analysis of the data presented in Fig.~\ref{fig:gapvis} 
and Tables~\ref{tab:znogap}-\ref{tab:tiogap}.
\begin{itemize}
\item Comparing the values in ($G_0W_0$, $\rho_\mathrm{val}$, $0$) 
row at (LDA, ES) and (GGA, ES) columns for ZnO, we find that 
the band gap varies by $3.21 - 2.82 = 0.39$ eV depending on 
the starting point (obtained from the LDA or GGA calculations). 
Averaging this quantity over different $\rho_\mathrm{core}=0$ 
rows for ZnO gives the mean variation in the band gap 
with different starting points as equal to 0.44 eV. 
Repeating this procedure for Cu$_2$O and TiO$_2$ 
yields the values of 0.06 eV and 0.04 eV, respectively. 
The large variation in the case of ZnO indicates that neither 
LDA nor GGA provides a good starting point for $GW$ calculations. 
At the same time, small variations for Cu$_2$O and TiO$_2$ 
imply that LDA and GGA give similar (but not necessarily good) 
starting points for $GW$ calculations. 
Other starting points were tried in $GW$ calculations 
for ZnO including DFT+U \cite{shih-10-zno-gw}, 
the screened hybrid functional \cite{schleife-08-zno-gw}, 
and the exact exchange optimized effective potential 
\cite{yan-11-zno-gw}. 
Note that the latter starting point can present some challenges 
in the subsequent $GW$ calculations \cite{fleszar-01-gw-oep}. 
Overall, the problem of the starting point in $GW$ calculations 
for ZnO may require further research to give the full picture.
\item Comparison of the values in $\rho_\mathrm{core}=0$ 
rows at (LDA, ES) and (LDA, TS) columns, as well as 
at (GGA, ES) and (GGA, TS) columns, shows that the 
variation of the band gap with the structural parameters is 
0.12 eV for ZnO, 0.10 eV for Cu$_2$O, and 0.09 eV for TiO$_2$. 
This suggests that the band gaps are fairly insensitive 
to the structural parameters.
\item Comparing the values in ($G_0W_0$, $\rho_\mathrm{val}$, $0$) 
and ($GW_0$, $\rho_\mathrm{val}$, $0$) rows, 
as well as in ($G_0W_0$, $\rho_\mathrm{scf}$, $0$) 
and ($GW_0$, $\rho_\mathrm{scf}$, $0$) rows, 
we find that the eigenvalue self-consistency 
in $G$ increases the band gap by 0.37 eV for ZnO, 
0.16 eV for Cu$_2$O, and 0.24 eV for TiO$_2$. This is 
consistent with previous studies \cite{shishkin-07-zno-gw}.
\item Comparison of the values in ($G_0W_0$, $\rho_\mathrm{val}$, $0$) 
and ($G_0W_0$, $\rho_\mathrm{scf}$, $0$) rows, 
as well as in ($GW_0$, $\rho_\mathrm{val}$, $0$) 
and ($GW_0$, $\rho_\mathrm{scf}$, $0$) rows, 
shows that the inclusion of core electrons in $\rho_\mathrm{ppm}$ 
increases the band gap of ZnO and Cu$_2$O by 0.51 and 0.15 eV, 
respectively, and decreases the band gap of TiO$_2$ by 0.22 eV. 
This is because for ZnO and Cu$_2$O, the VBM is lowered 
by a larger amount than the CBM, while the opposite 
scenario takes place for TiO$_2$, as follows from 
Supplemental Material \cite{nlcc_suppl}.
\item Comparing the values in $\rho_\mathrm{core}=0$ and 
$\rho_\mathrm{core}\neq0$ rows, we find that the inclusion of NLCC 
in the matrix elements of $V_\mathrm{xc}$ decreases the band gap 
of ZnO and Cu$_2$O by 0.70 and 0.69 eV, respectively, and increases 
the band gap of TiO$_2$ by 0.27 eV. This is due to the fact that for 
ZnO and Cu$_2$O, the VBM is raised by a larger amount than the CBM, 
while the opposite holds for TiO$_2$, as one can see from 
Supplemental Material \cite{nlcc_suppl}.
\end{itemize}
Overall, the largest variation of the band gap comes from 
the inclusion of NLCC in the matrix elements of $V_\mathrm{xc}$. 
This inclusion introduces significant errors in the calculated 
band gaps.

Fair agreement is found when comparing our results to those 
of previous $GW$ calculations for each oxide and specific flavor. 
In line with the criticism of the HL PPM 
\cite{stankovski-11-zno-gw, miglio-12-zno-gw, larson-13-zno-ppm}, 
previous HL PPM calculations are compared 
to our $\rho_\mathrm{ppm}=\rho_\mathrm{scf}$ results, 
while previous GN PPM and non-PPM calculations 
to our $\rho_\mathrm{ppm}=\rho_\mathrm{val}$ results.
\begin{itemize}
\item For ZnO, we focus on (LDA, ES) column in 
Fig.~\ref{fig:gapvis}(a) or Table~\ref{tab:znogap}. 
The most accurate non-PPM $G_0W_0$ calculations 
gave the following values for the band gap: 
2.83 eV with the full potential linearized augmented 
plane wave (FLAPW) method \cite{friedrich-11-zno-gw} 
and 2.87 eV with the projector augmented wave 
(PAW) method \cite{klimes-14-zno-gw}. 
We find that the HL PPM gives a somewhat larger value of 3.21 eV 
and a substantially larger value of 3.82 eV when $\rho_\mathrm{ppm}$ 
is set to $\rho_\mathrm{val}$ and $\rho_\mathrm{scf}$, respectively. 
Previous HL PPM $G_0W_0$ calculations showed values in this range, 
3.4 eV \cite{shih-10-zno-gw}, 3.57 eV \cite{stankovski-11-zno-gw}, 
and 3.56 eV \cite{miglio-12-zno-gw}, 
with one exception where the value of 
2.80 eV was reported \cite{larson-13-zno-ppm}. 
Other studies reported much lower values, 
such as non-PPM $G_0W_0$ band gaps of 2.17--2.43 eV 
\cite{miglio-12-zno-gw, larson-13-zno-ppm} and 
the GN PPM $G_0W_0$ band gaps of 2.27--2.56 eV 
\cite{stankovski-11-zno-gw, miglio-12-zno-gw, 
berger-12-zno-gw-eet, larson-13-zno-ppm}. 
We do not compare to the results of 
Refs.~\cite{shishkin-07-zno-gw, fuchs-07-zno-gw, 
kotani-07-zno-scgw, bechstedt-09-scgw, huser-13-zno-gw} 
which may be affected by the basis set convergence issues 
as discussed in Ref.~\onlinecite{klimes-14-zno-gw}.
\item For Cu$_2$O, let us look at (LDA, ES) column in 
Fig.~\ref{fig:gapvis}(b) or Table~\ref{tab:cuogap}. 
The ($G_0W_0$, $\rho_\mathrm{val}$, $0$) 
band gap of 1.56 eV compares with non-PPM $G_0W_0$ 
value of 1.34 eV \cite{bruneval-06-cu2o-scgw}. 
The ($GW_0$, $\rho_\mathrm{val}$, $0$) 
band gap of 1.77 eV is close to 
non-PPM eigenvalue self-consistent $GW$ 
band gap of 1.80 eV \cite{bruneval-06-cu2o-scgw}.
\item For TiO$_2$, we start with (LDA, ES) column in 
Fig.~\ref{fig:gapvis}(c) or Table~\ref{tab:tiogap}. 
The ($G_0W_0$, $\rho_\mathrm{val}$, $0$) 
band gap of 3.44 eV is close to non-PPM $G_0W_0$ 
value of 3.34 eV \cite{kang-10-tio2-gw}. 
We now move on to (GGA, TS) column. 
The ($G_0W_0$, $\rho_\mathrm{val}$, $0$) 
band gap of 3.34 eV is comparable to the GN PPM $G_0W_0$ 
value of 3.59 eV \cite{chiodo-10-tio2-gw}. 
The ($G_0W_0$, $\rho_\mathrm{scf}$, $0$) 
band gap of 3.14 eV is close to the HL PPM $G_0W_0$ 
value of 3.13 eV \cite{malashevich-14-tio2-gw}.
\end{itemize}

We now compare the calculated band gaps with the experimental data. 
The absolute difference of the experimental band gap and 
the value in ($G_0W_0$, $\rho_\mathrm{val}$, $0$) row 
at (LDA, ES) column for ZnO is equal to 0.23 eV. 
Averaging this quantity over the four columns in Fig.~\ref{fig:gapvis}(a) 
or Table~\ref{tab:znogap} gives the value of 0.43 eV. Further averaging 
over the three TMOs gives the mean deviation from experiment of 0.40 eV. 
This procedure is repeated for each flavor represented by different row 
in Fig.~\ref{fig:gapvis} and Tables~\ref{tab:znogap}-\ref{tab:tiogap}. 
Among the eight flavors, the smallest deviation of 0.18 eV is found 
for ($GW_0$, $\rho_\mathrm{val}$, $0$) flavor, 
followed by the 0.30 eV deviation 
for ($GW_0$, $\rho_\mathrm{scf}$, $0$) flavor, 
the 0.31 eV deviation 
for ($G_0W_0$, $\rho_\mathrm{scf}$, $0$) flavor, 
and the 0.40 eV deviation 
for ($G_0W_0$, $\rho_\mathrm{val}$, $0$) flavor. 
Corresponding deviations for $\rho_\mathrm{core}\neq0$ rows 
fall within the 0.49 to 0.78 eV range. We conclude that 
the best overall agreement with experiment is obtained for 
($GW_0$, $\rho_\mathrm{val}$, $0$) flavor. 
On the other hand, we note from Fig.~\ref{fig:gapvis} 
that ($GW_0$, $\rho_\mathrm{val}$, $0$) values irregularly 
underestimate and overestimate the experimental band gaps, 
while ($G_0W_0$, $\rho_\mathrm{val}$, $0$) values always 
underestimate the experiment, suggesting that 
the latter flavor may be preferable to the former. 
Yet this conclusion may be deceiving given that 
the HL PPM with $\rho_\mathrm{val}$ overestimates 
the non-PPM band gap of ZnO by 0.34--0.38 eV 
(see the comparison with the previous calculations above) 
and that the experimental band gaps are renormalized by 
electron-phonon interaction not included in our calculations. 
Overall, it may be premature to conclude which flavor is 
preferable for TMOs until the effect of the vertex correction 
on $GW$ band gaps of these materials is thoroughly studied.

\section{Summary}
\label{sec:sum}

In summary, we quantify the effects of different approximations used 
in the $GW$ method on the band gaps and band edges for three TMOs: 
wurtzite ZnO, cuprite Cu$_2$O, and rutile TiO$_2$. It is found that 
the $GW$ band gap of ZnO is sensitive to the starting point 
obtained from the LDA or GGA calculations, 
suggesting that the Kohn-Sham orbitals 
differ from the quasiparticle orbitals. 
It is shown that the HL PPM becomes similar 
to the GN PPM and gives better agreement with the RPA 
when $\rho_\mathrm{ppm}$ is set to $\rho_\mathrm{val}$, 
that is, only the valence electrons are used to determine 
the effective bare plasma frequencies for the HL PPM. 
It is demonstrated that the theoretically justified choice 
of approximations, namely eigenvalue self-consistent $GW_0$ 
scheme, $\rho_\mathrm{val}$ in the HL PPM, and the proper 
treatment of the $V_\mathrm{xc}$ term, give the best overall 
agreement between the calculated and measured band gaps.

\begin{acknowledgments}
We are grateful to Dr. Brad D. Malone for implementing 
the $\rho_\mathrm{ppm}=\rho_\mathrm{val}$ option in 
the \texttt{Quantum ESPRESSO} interface to \texttt{BerkeleyGW}. 
We thank Mr. Felipe H. Jornada for implementing the Delaunay 
tessellation for band structure interpolation in \texttt{BerkeleyGW}. 
We acknowledge helpful comments on an early version of the paper by 
Prof. Peihong Zhang, Prof. Gian-Marco Rignanese, and Mr. Derek Vigil. 
G.S. and B.K. acknowledge support from DOE (Grant No. $\mathrm{DE}$-$\mathrm{EE0004840}$) 
and NSF (Grant No. 1048796), C.H.P. from Korean NRF funded by MSIP 
(Grant No. NRF-2013R1A1A1076141). 
This research used resources of the Oak Ridge Leadership Computing 
Facility located in the Oak Ridge National Laboratory, which is 
supported by the Office of Science of the Department of Energy 
under Contract DE-AC05-00OR22725.
\end{acknowledgments}

\end{document}

\title{Supplemental material: Insights and challenges of applying the $GW$ method to transition metal oxides}

\author{Georgy Samsonidze}
\affiliation{Research and Technology Center, Robert Bosch LLC, Cambridge, Massachusetts 02142, USA}
\author{Cheol-Hwan Park}
\affiliation{Department of Physics, Seoul National University, Seoul 151-747, Korea}
\author{Boris Kozinsky}
\affiliation{Research and Technology Center, Robert Bosch LLC, Cambridge, Massachusetts 02142, USA}

\date{\today}

\maketitle

\begin{table*}
\caption{Kohn-Sham and quasiparticle energies ($E^\mathrm{KS}$ and $E^\mathrm{QP}$) 
and matrix elements of $V_\mathrm{xc}$ and $\Sigma$ 
at the valence band maximum (VBM) and conduction band minimum (CBM) 
of wurtzite ZnO calculated within DFT and $GW$ 
using different exchange-correlation functionals (LDA and GGA), 
experimental and theoretical structural parameters (ES and TS), 
non-self-consistent $G_0W_0$ and eigenvalue self-consistent $GW_0$ schemes, 
HL PPM with $\rho_\mathrm{ppm} = \rho_\mathrm{val}$ and $\rho_\mathrm{scf}$, 
and matrix elements of $V_\mathrm{xc}$ without and with NLCC 
($\rho_\mathrm{core} = 0$ and $\neq 0$). 
The VBM and CBM are at the $\Gamma$ point. 
All values are in eV. 
The zero reference for the energy scale is 
the average ($\mathbf{G} = \mathbf{0}$ component) 
electrostatic (ionic plus Hartree) potential.
\label{tab:znoedge}}
\begin{ruledtabular}
\begin{tabular}{llllrrrrrrrr}
\multicolumn{4}{l}{Wurtzite ZnO}                                                                                               & \multicolumn{4}{c}{LDA}                                                       & \multicolumn{4}{c}{GGA}                                                       \\
~                         & ~                                    & ~                        &                                  & \multicolumn{2}{c}{ES}                & \multicolumn{2}{c}{TS}                & \multicolumn{2}{c}{ES}                & \multicolumn{2}{c}{TS}                \\
~                         & $\rho_\mathrm{ppm}$                  & $\rho_\mathrm{core}$     &                                  & VBM               & CBM               & VBM               & CBM               & VBM               & CBM               & VBM               & CBM               \\
\hline
\multicolumn{2}{l}{\multirow{3}{*}{DFT}}                         & ~                        & $E^\mathrm{KS}$                  &         $  7.88$  &         $  8.63$  &         $  8.62$  &         $  9.43$  &         $  8.34$  &         $  9.19$  &         $  7.72$  &         $  8.50$  \\
                          &                                      & $=0$                     & \multirow{2}{*}{$V_\mathrm{xc}$} &         $-26.66$  &         $-13.35$  &         $-27.13$  &         $-13.51$  &         $-27.06$  &         $-12.77$  &         $-26.52$  &         $-12.58$  \\
                          &                                      & $\neq0$                  &                                  &         $-28.02$  &         $-13.76$  &         $-28.54$  &         $-13.92$  &         $-27.98$  &         $-13.04$  &         $-27.42$  &         $-12.85$  \\
\multirow{8}{*}{$G_0W_0$} & \multirow{4}{*}{$\rho_\mathrm{val}$} & \multirow{2}{*}{$=0$}    & $E^\mathrm{QP}$                  &         $  6.00$  &         $  9.21$  &         $  6.70$  &         $ 10.02$  &         $  6.63$  &         $  9.45$  &         $  6.05$  &         $  8.74$  \\
                          &                                      &                          & $\Sigma$                         &         $-28.54$  &         $-12.77$  &         $-29.06$  &         $-12.91$  &         $-28.77$  &         $-12.52$  &         $-28.20$  &         $-12.34$  \\
                          &                                      & \multirow{2}{*}{$\neq0$} & $E^\mathrm{QP}$                  &         $  7.04$  &         $  9.54$  &         $  7.77$  &         $ 10.36$  &         $  7.34$  &         $  9.67$  &         $  6.73$  &         $  8.96$  \\
                          &                                      &                          & $\Sigma$                         &         $-28.86$  &         $-12.85$  &         $-29.39$  &         $-12.98$  &         $-28.98$  &         $-12.56$  &         $-28.41$  &         $-12.39$  \\
                          & \multirow{4}{*}{$\rho_\mathrm{scf}$} & \multirow{2}{*}{$=0$}    & $E^\mathrm{QP}$                  &         $  4.85$  &         $  8.67$  &         $  5.55$  &         $  9.48$  &         $  5.51$  &         $  8.89$  &         $  4.93$  &         $  8.19$  \\
                          &                                      &                          & $\Sigma$                         &         $-29.69$  &         $-13.31$  &         $-30.21$  &         $-13.45$  &         $-29.88$  &         $-13.07$  &         $-29.31$  &         $-12.89$  \\
                          &                                      & \multirow{2}{*}{$\neq0$} & $E^\mathrm{QP}$                  &         $  6.03$  &         $  9.04$  &         $  6.77$  &         $  9.86$  &         $  6.32$  &         $  9.14$  &         $  5.71$  &         $  8.43$  \\
                          &                                      &                          & $\Sigma$                         &         $-29.87$  &         $-13.35$  &         $-30.40$  &         $-13.49$  &         $-30.01$  &         $-13.10$  &         $-29.43$  &         $-12.92$  \\
\multirow{8}{*}{$GW_0$}   & \multirow{4}{*}{$\rho_\mathrm{val}$} & \multirow{2}{*}{$=0$}    & $E^\mathrm{QP}$                  &         $  5.53$  &         $  9.21$  &         $  6.22$  &         $ 10.03$  &         $  6.16$  &         $  9.41$  &         $  5.58$  &         $  8.70$  \\
                          &                                      &                          & $\Sigma$                         &         $-29.01$  &         $-12.77$  &         $-29.54$  &         $-12.90$  &         $-29.23$  &         $-12.56$  &         $-28.66$  &         $-12.39$  \\
                          &                                      & \multirow{2}{*}{$\neq0$} & $E^\mathrm{QP}$                  &         $  6.85$  &         $  9.66$  &         $  7.58$  &         $ 10.49$  &         $  7.07$  &         $  9.71$  &         $  6.46$  &         $  9.00$  \\
                          &                                      &                          & $\Sigma$                         &         $-29.05$  &         $-12.73$  &         $-29.58$  &         $-12.86$  &         $-29.25$  &         $-12.52$  &         $-28.68$  &         $-12.35$  \\
                          & \multirow{4}{*}{$\rho_\mathrm{scf}$} & \multirow{2}{*}{$=0$}    & $E^\mathrm{QP}$                  &         $  4.48$  &         $  8.61$  &         $  5.18$  &         $  9.43$  &         $  5.15$  &         $  8.81$  &         $  4.57$  &         $  8.11$  \\
                          &                                      &                          & $\Sigma$                         &         $-30.06$  &         $-13.37$  &         $-30.58$  &         $-13.50$  &         $-30.25$  &         $-13.15$  &         $-29.67$  &         $-12.98$  \\
                          &                                      & \multirow{2}{*}{$\neq0$} & $E^\mathrm{QP}$                  &         $  5.82$  &         $  9.05$  &         $  6.56$  &         $  9.87$  &         $  6.06$  &         $  9.10$  &         $  5.45$  &         $  8.39$  \\
                          &                                      &                          & $\Sigma$                         &         $-30.08$  &         $-13.34$  &         $-30.60$  &         $-13.48$  &         $-30.26$  &         $-13.13$  &         $-29.69$  &         $-12.96$  \\
\end{tabular}
\end{ruledtabular}
\end{table*}

\begin{table*}
\caption{Kohn-Sham and quasiparticle energies ($E^\mathrm{KS}$ and $E^\mathrm{QP}$) 
and matrix elements of $V_\mathrm{xc}$ and $\Sigma$ 
at the valence band maximum (VBM) and conduction band minimum (CBM) 
of cuprite Cu$_2$O calculated within DFT and $GW$ 
using different exchange-correlation functionals (LDA and GGA), 
experimental and theoretical structural parameters (ES and TS), 
non-self-consistent $G_0W_0$ and eigenvalue self-consistent $GW_0$ schemes, 
HL PPM with $\rho_\mathrm{ppm} = \rho_\mathrm{val}$ and $\rho_\mathrm{scf}$, 
and matrix elements of $V_\mathrm{xc}$ without and with NLCC 
($\rho_\mathrm{core} = 0$ and $\neq 0$). 
The VBM and CBM are at the $\Gamma$ point. 
All values are in eV. 
The zero reference for the energy scale is 
the average ($\mathbf{G} = \mathbf{0}$ component) 
electrostatic (ionic plus Hartree) potential.
\label{tab:cuoedge}}
\begin{ruledtabular}
\begin{tabular}{llllrrrrrrrr}
\multicolumn{4}{l}{Cuprite Cu$_2$O}                                                                                            & \multicolumn{4}{c}{LDA}                                                       & \multicolumn{4}{c}{GGA}                                                       \\
~                         & ~                                    & ~                        &                                  & \multicolumn{2}{c}{ES}                & \multicolumn{2}{c}{TS}                & \multicolumn{2}{c}{ES}                & \multicolumn{2}{c}{TS}                \\
~                         & $\rho_\mathrm{ppm}$                  & $\rho_\mathrm{core}$     &                                  & VBM               & CBM               & VBM               & CBM               & VBM               & CBM               & VBM               & CBM               \\
\hline
\multicolumn{2}{l}{\multirow{3}{*}{DFT}}                         & ~                        & $E^\mathrm{KS}$                  &         $  9.52$  &         $ 10.04$  &         $ 10.67$  &         $ 11.36$  &         $  9.99$  &         $ 10.52$  &         $  9.60$  &         $ 10.07$  \\
                          &                                      & $=0$                     & \multirow{2}{*}{$V_\mathrm{xc}$} &         $-32.10$  &         $-30.69$  &         $-32.24$  &         $-30.86$  &         $-32.40$  &         $-30.95$  &         $-32.35$  &         $-30.88$  \\
                          &                                      & $\neq0$                  &                                  &         $-33.96$  &         $-32.49$  &         $-34.13$  &         $-32.68$  &         $-33.79$  &         $-32.29$  &         $-33.73$  &         $-32.21$  \\
\multirow{8}{*}{$G_0W_0$} & \multirow{4}{*}{$\rho_\mathrm{val}$} & \multirow{2}{*}{$=0$}    & $E^\mathrm{QP}$                  &         $  8.73$  &         $ 10.29$  &         $  9.89$  &         $ 11.60$  &         $  9.37$  &         $ 10.88$  &         $  8.97$  &         $ 10.43$  \\
                          &                                      &                          & $\Sigma$                         &         $-32.89$  &         $-30.44$  &         $-33.02$  &         $-30.62$  &         $-33.03$  &         $-30.58$  &         $-32.98$  &         $-30.52$  \\
                          &                                      & \multirow{2}{*}{$\neq0$} & $E^\mathrm{QP}$                  &         $ 10.04$  &         $ 11.18$  &         $ 11.22$  &         $ 12.09$  &         $ 10.34$  &         $ 11.25$  &         $  9.93$  &         $ 10.94$  \\
                          &                                      &                          & $\Sigma$                         &         $-33.44$  &         $-31.35$  &         $-33.57$  &         $-31.95$  &         $-33.44$  &         $-31.56$  &         $-33.39$  &         $-31.35$  \\
                          & \multirow{4}{*}{$\rho_\mathrm{scf}$} & \multirow{2}{*}{$=0$}    & $E^\mathrm{QP}$                  &         $  7.67$  &         $  9.43$  &         $  8.83$  &         $ 10.75$  &         $  8.35$  &         $ 10.05$  &         $  7.94$  &         $  9.59$  \\
                          &                                      &                          & $\Sigma$                         &         $-33.95$  &         $-31.30$  &         $-34.07$  &         $-31.47$  &         $-34.05$  &         $-31.42$  &         $-34.00$  &         $-31.36$  \\
                          &                                      & \multirow{2}{*}{$\neq0$} & $E^\mathrm{QP}$                  &         $  9.24$  &         $ 10.50$  &         $ 10.43$  &         $ 11.39$  &         $  9.51$  &         $ 10.52$  &         $  9.10$  &         $ 10.22$  \\
                          &                                      &                          & $\Sigma$                         &         $-34.24$  &         $-32.03$  &         $-34.37$  &         $-32.65$  &         $-34.27$  &         $-32.28$  &         $-34.22$  &         $-32.07$  \\
\multirow{8}{*}{$GW_0$}   & \multirow{4}{*}{$\rho_\mathrm{val}$} & \multirow{2}{*}{$=0$}    & $E^\mathrm{QP}$                  &         $  8.46$  &         $ 10.23$  &         $  9.63$  &         $ 11.55$  &         $  9.12$  &         $ 10.83$  &         $  8.72$  &         $ 10.37$  \\
                          &                                      &                          & $\Sigma$                         &         $-33.16$  &         $-30.50$  &         $-33.28$  &         $-30.67$  &         $-33.27$  &         $-30.64$  &         $-33.23$  &         $-30.57$  \\
                          &                                      & \multirow{2}{*}{$\neq0$} & $E^\mathrm{QP}$                  &         $ 10.21$  &         $ 11.34$  &         $ 11.40$  &         $ 12.25$  &         $ 10.42$  &         $ 11.32$  &         $ 10.01$  &         $ 11.01$  \\
                          &                                      &                          & $\Sigma$                         &         $-33.27$  &         $-31.19$  &         $-33.39$  &         $-31.79$  &         $-33.35$  &         $-31.48$  &         $-33.31$  &         $-31.28$  \\
                          & \multirow{4}{*}{$\rho_\mathrm{scf}$} & \multirow{2}{*}{$=0$}    & $E^\mathrm{QP}$                  &         $  7.37$  &         $  9.24$  &         $  8.54$  &         $ 10.57$  &         $  8.06$  &         $  9.86$  &         $  7.65$  &         $  9.40$  \\
                          &                                      &                          & $\Sigma$                         &         $-34.25$  &         $-31.49$  &         $-34.36$  &         $-31.65$  &         $-34.33$  &         $-31.60$  &         $-34.29$  &         $-31.54$  \\
                          &                                      & \multirow{2}{*}{$\neq0$} & $E^\mathrm{QP}$                  &         $  9.17$  &         $ 10.50$  &         $ 10.36$  &         $ 11.39$  &         $  9.39$  &         $ 10.47$  &         $  8.98$  &         $ 10.16$  \\
                          &                                      &                          & $\Sigma$                         &         $-34.31$  &         $-32.03$  &         $-34.43$  &         $-32.65$  &         $-34.38$  &         $-32.33$  &         $-34.34$  &         $-32.12$  \\
\end{tabular}
\end{ruledtabular}
\end{table*}

\begin{table*}
\caption{Kohn-Sham and quasiparticle energies ($E^\mathrm{KS}$ and $E^\mathrm{QP}$) 
and matrix elements of $V_\mathrm{xc}$ and $\Sigma$ 
at the valence band maximum (VBM) and conduction band minimum (CBM) 
of rutile TiO$_2$ calculated within DFT and $GW$ 
using different exchange-correlation functionals (LDA and GGA), 
experimental and theoretical structural parameters (ES and TS), 
non-self-consistent $G_0W_0$ and eigenvalue self-consistent $GW_0$ schemes, 
HL PPM with $\rho_\mathrm{ppm} = \rho_\mathrm{val}$ and $\rho_\mathrm{scf}$, 
and matrix elements of $V_\mathrm{xc}$ without and with NLCC 
($\rho_\mathrm{core} = 0$ and $\neq 0$). 
The VBM and CBM are at the $\Gamma$ point (in regular font) 
and at the R point (in cursive font). 
All values are in eV. 
The zero reference for the energy scale is 
the average ($\mathbf{G} = \mathbf{0}$ component) 
electrostatic (ionic plus Hartree) potential.
\label{tab:tioedge}}
\begin{ruledtabular}
\begin{tabular}{llllrrrrrrrr}
\multicolumn{4}{l}{Rutile TiO$_2$}                                                                                             & \multicolumn{4}{c}{LDA}                                                       & \multicolumn{4}{c}{GGA}                                                       \\
~                         & ~                                    & ~                        &                                  & \multicolumn{2}{c}{ES}                & \multicolumn{2}{c}{TS}                & \multicolumn{2}{c}{ES}                & \multicolumn{2}{c}{TS}                \\
~                         & $\rho_\mathrm{ppm}$                  & $\rho_\mathrm{core}$     &                                  & VBM               & CBM               & VBM               & CBM               & VBM               & CBM               & VBM               & CBM               \\
\hline
\multicolumn{2}{l}{\multirow{3}{*}{DFT}}                         & ~                        & $E^\mathrm{KS}$                  &         $  8.07$  &         $  9.88$  &         $  8.45$  &         $ 10.30$  &         $  8.43$  &         $ 10.33$  &         $  8.00$  &         $  9.84$  \\
                          &                                      & $=0$                     & \multirow{2}{*}{$V_\mathrm{xc}$} &         $-20.29$  &         $-20.31$  &         $-20.36$  &         $-20.33$  &         $-20.56$  &         $-20.36$  &         $-20.48$  &         $-20.36$  \\
                          &                                      & $\neq0$                  &                                  &         $-21.12$  &         $-21.44$  &         $-21.19$  &         $-21.45$  &         $-21.04$  &         $-21.14$  &         $-20.97$  &         $-21.14$  \\
\multirow{8}{*}{$G_0W_0$} & \multirow{4}{*}{$\rho_\mathrm{val}$} & \multirow{2}{*}{$=0$}    & $E^\mathrm{QP}$                  &         $  6.90$  & $\mathit{ 10.34}$ &         $  7.27$  & $\mathit{ 10.80}$ &         $  7.42$  & $\mathit{ 10.84}$ &         $  7.00$  & $\mathit{ 10.35}$ \\
                          &                                      &                          & $\Sigma$                         &         $-21.46$  & $\mathit{-19.38}$ &         $-21.54$  & $\mathit{-19.35}$ &         $-21.57$  & $\mathit{-19.39}$ &         $-21.48$  & $\mathit{-19.38}$ \\
                          &                                      & \multirow{2}{*}{$\neq0$} & $E^\mathrm{QP}$                  &         $  7.55$  & $\mathit{ 11.23}$ &         $  7.92$  & $\mathit{ 11.69}$ &         $  7.79$  & $\mathit{ 11.44}$ &         $  7.38$  & $\mathit{ 10.95}$ \\
                          &                                      &                          & $\Sigma$                         &         $-21.64$  & $\mathit{-19.58}$ &         $-21.73$  & $\mathit{-19.54}$ &         $-21.68$  & $\mathit{-19.54}$ &         $-21.59$  & $\mathit{-19.53}$ \\
                          & \multirow{4}{*}{$\rho_\mathrm{scf}$} & \multirow{2}{*}{$=0$}    & $E^\mathrm{QP}$                  &         $  6.58$  & $\mathit{  9.87}$ &         $  6.96$  & $\mathit{ 10.31}$ &         $  7.11$  & $\mathit{ 10.34}$ &         $  6.69$  &         $  9.84$  \\
                          &                                      &                          & $\Sigma$                         &         $-21.77$  & $\mathit{-19.86}$ &         $-21.85$  & $\mathit{-19.83}$ &         $-21.88$  & $\mathit{-19.89}$ &         $-21.79$  &         $-20.36$  \\
                          &                                      & \multirow{2}{*}{$\neq0$} & $E^\mathrm{QP}$                  &         $  7.26$  & $\mathit{ 10.83}$ &         $  7.64$  & $\mathit{ 11.27}$ &         $  7.51$  & $\mathit{ 10.99}$ &         $  7.09$  &         $ 10.50$  \\
                          &                                      &                          & $\Sigma$                         &         $-21.93$  & $\mathit{-19.98}$ &         $-22.00$  & $\mathit{-19.96}$ &         $-21.97$  & $\mathit{-19.99}$ &         $-21.88$  &         $-20.48$  \\
\multirow{8}{*}{$GW_0$}   & \multirow{4}{*}{$\rho_\mathrm{val}$} & \multirow{2}{*}{$=0$}    & $E^\mathrm{QP}$                  &         $  6.66$  & $\mathit{ 10.39}$ &         $  7.03$  & $\mathit{ 10.85}$ &         $  7.19$  & $\mathit{ 10.88}$ &         $  6.77$  & $\mathit{ 10.39}$ \\
                          &                                      &                          & $\Sigma$                         &         $-21.69$  & $\mathit{-19.33}$ &         $-21.78$  & $\mathit{-19.30}$ &         $-21.80$  & $\mathit{-19.34}$ &         $-21.71$  & $\mathit{-19.33}$ \\
                          &                                      & \multirow{2}{*}{$\neq0$} & $E^\mathrm{QP}$                  &         $  7.48$  & $\mathit{ 11.51}$ &         $  7.85$  & $\mathit{ 11.97}$ &         $  7.67$  & $\mathit{ 11.65}$ &         $  7.26$  & $\mathit{ 11.15}$ \\
                          &                                      &                          & $\Sigma$                         &         $-21.71$  & $\mathit{-19.30}$ &         $-21.79$  & $\mathit{-19.26}$ &         $-21.80$  & $\mathit{-19.33}$ &         $-21.71$  & $\mathit{-19.33}$ \\
                          & \multirow{4}{*}{$\rho_\mathrm{scf}$} & \multirow{2}{*}{$=0$}    & $E^\mathrm{QP}$                  &         $  6.33$  & $\mathit{  9.81}$ &         $  6.70$  & $\mathit{ 10.26}$ &         $  6.86$  & $\mathit{ 10.29}$ &         $  6.44$  &         $  9.80$  \\
                          &                                      &                          & $\Sigma$                         &         $-22.03$  & $\mathit{-19.91}$ &         $-22.10$  & $\mathit{-19.89}$ &         $-22.13$  & $\mathit{-19.93}$ &         $-22.04$  &         $-20.40$  \\
                          &                                      & \multirow{2}{*}{$\neq0$} & $E^\mathrm{QP}$                  &         $  7.15$  & $\mathit{ 10.94}$ &         $  7.53$  & $\mathit{ 11.38}$ &         $  7.34$  & $\mathit{ 11.05}$ &         $  6.93$  & $\mathit{ 10.56}$ \\
                          &                                      &                          & $\Sigma$                         &         $-22.04$  & $\mathit{-19.87}$ &         $-22.12$  & $\mathit{-19.85}$ &         $-22.13$  & $\mathit{-19.93}$ &         $-22.04$  & $\mathit{-19.92}$ \\
\end{tabular}
\end{ruledtabular}
\end{table*}